\documentclass[final,5p,times,twocolumn]{elsarticle}
\usepackage{graphicx}
\usepackage{lineno}
\usepackage{siunitx}
\usepackage{amsmath}
\usepackage{comment}
\usepackage{color}

\begin{document}

\begin{frontmatter}

\title{Low-Noise SiPM Light Readout and ASIC-Based Charge Readout of a Liquid Argon Time Projection Chamber for MeV Gamma-Ray Measurements}

\author[a,b]{Satoshi Takashima \corref{auth}}
\author[b,c]{Hirokazu Odaka}
\author[d]{Shota Arai}
\author[b]{Kentaro Shirahama}
\author[b]{Ryutaro Tatsumi}
\author[b]{Hodaka Kawamura}
\author[b]{\\Masashi Tanaka}
\author[e]{Shintaro Arai}
\author[f]{Tsuguo Aramaki}
\author[d,g,h]{Aya Bamba}
\author[i]{Lorenzo Fabris}
\author[j]{Georgia Karagiorgi}
\author[b]{\\Haruki Kuramoto} 
\author[f]{Jonathan LeyVa}
\author[k]{John W. Mitchell}
\author[b]{Aiko Miyamoto}
\author[l]{Reshmi Mukherjee} 
\author[b]{Kaito Murakami}
\author[b]{\\Azuki Nagao}
\author[f]{Arathi Suraj}
\author[b]{Sayana Takatsuka}
\author[e]{Chihiro Watanabe}
\author[m]{Shin Watanabe} 
\author[e]{Yutaro Yano}
\author[n]{\\Kazushi Yawata}
\author[o]{Hiroki Yoneda}
\author[d]{Kohei Yorita}
\author[f]{Jiancheng Zeng}

\cortext[auth]{ Corresponding author. e-mail: satoshi.takashima@riken.jp}

\address[a]{RIKEN Nishina Center, 2-1 Hirosawa, Wako, Saitama 351-0198, Japan}
\address[b]{Department of Earth and Space Science, University of Osaka, 1-1 Machikaneyama, Toyonaka, Osaka, 560-0043, Japan}
\address[c]{Kavli IPMU, University of Tokyo, 5-1-5 Kashiwanoha, Kashiwa, Chiba, 277-8583, Japan}
\address[d]{Department of Physics, University of Tokyo, 7-3-1 Hongo, Bunkyo-ku, Tokyo, 113-0033, Japan}
\address[e]{Department of Physics, Waseda University, 3-4-1 Okubo, Shinjuku-ku, Tokyo, 169-8555, Japan}
\address[f]{Department of Physics, Northeastern University, 360 Huntington Avenue, Boston, MA, 02115, USA}
\address[g]{Research Center for the Early Universe, Faculty of Science, University of Tokyo, 7-3-1 Hongo, Bunkyo-ku, Tokyo, 113-0033, Japan}
\address[h]{Trans-Scale Quantum Science Institute, University of Tokyo, 7-3-1 Hongo, Bunkyo-ku, Tokyo, 113-0033, Japan}
\address[i]{Oak ridge National Laboratory, 5200, 1 Bethel Valley Rd, Oark Ridge, TN, 37830, USA}
\address[j]{Department of Physics, Columbia University, New York, NY, 10027, USA}
\address[k]{NASA Goddard Space Flight Center, 8800 Greenbelt Road, Greenbelt, MD, 20771, USA}
\address[l]{Department of Physics and Astronomy, Barnard College, 3009 Broadway, New York, NY, 10027, USA}
\address[m]{Japan Aerospace Exploration Agency (JAXA), 3-1-1 Yoshinodai, Chuo-ku, Sagamihara, Kanagawa, 169-8555, Japan}
\address[n]{National Defense Medical College, 3-2 Namiki, Tokorozawa, Saitama, 359-8513, Japan}
\address[o]{Department of Physics, Kyoto University, Kitashirakawa Oiwakecho, Sakyo-ku, Kyoto, Kyoto, 606-8502, Japan}

\begin{abstract}
We have developed a compact liquid argon time projection chamber (LArTPC), NanoGRAMS, as a technology demonstrator for the Gamma-Ray and AntiMatter Survey (GRAMS). 
LArTPCs have the potential to enable Compton cameras with unprecedented effective area in the MeV gamma-ray band.
NanoGRAMS has an active volume of $5.12 \times 5.12 \times 10~\mathrm{cm^3}$ and is equipped with a low-noise scintillation and charge readout system. 
The scintillation light is detected by an array of 16 SiPMs ($6 \times 6~\mathrm{mm^2}$ each), whose signals are summed and amplified by a low-noise transimpedance amplifier operable at liquid argon temperature. 
Ionization electrons are read out with \SI{3.2}{mm}-pitch pixels and processed by VATA-SGD ASICs, with synchronization provided by an FPGA-based data acquisition system.
We irradiated the detector with a $^{60}\mathrm{Co}$ source (1173 and \SI{1332}{keV}) and successfully detected both 1-hit and 2-hit events. 
The collected charge was converted to deposited energy using a phenomenological recombination model, and the detector response was evaluated with a Geant4-based Monte Carlo simulation. 
The reconstructed energy spectrum shows Compton edges at 963 and \SI{1118}{keV}, consistent with the expected values. 
For 2-hit events, the sequence of interactions was identified, and the reconstructed back-projection image agrees with the source position. 
These results demonstrate the feasibility of NanoGRAMS as a Compton camera for MeV gamma-ray imaging spectroscopy.


\end{abstract}
\end{frontmatter}

\section{Introduction}
\label{sec:intro}
MeV gamma rays, covering energies from a few hundred keV to a few tens of MeV, represent the last unexplored window in multiwavelength astronomy.
This energy range is not only characteristic of nuclear de-excitation gamma rays but also forms a transition region between thermal and non-thermal emission, making it crucial for understanding particle acceleration mechanisms in astrophysical environments.
In the MeV energy band, the Compton camera is a promising technique for imaging spectroscopy, as Compton scattering is the dominant interaction process of gamma rays.
As a space-based Compton camera, COMPTEL aboard the Compton Gamma-Ray Observatory (1991--2000) played a pioneering role and identified more than 30 steady gamma-ray sources \citep{Schoenfelder1993}.
More recently, COSI-SMEX has been selected as a space mission to deploy a germanium-based Compton camera, featuring excellent energy resolution with a planned launch in 2027 \citep{Tomsick2024}.

MeV gamma-ray astronomy is technically challenging, mainly for three reasons.
First, photons in the MeV energy band have low interaction rates in detector materials compared with lower-energy X-rays and higher-energy GeV gamma rays.
Second, focusing optics for gamma rays are extremely difficult to realize.
Third, intense background particles in orbit can contaminate the measurements.
These challenges have contributed to the limited number of MeV gamma-ray missions and the 27-year observational gap between COMPTEL and COSI-SMEX.

The Gamma-Ray and AntiMatter Survey (GRAMS) is a proposed balloon-borne/satellite mission that utilizes a liquid argon time projection chamber (LArTPC) as a Compton camera \citep{Aramaki2020}.
LArTPCs are three-dimensional detectors capable of measuring both the energy loss and the spatial positions of particle interactions, and the detector technology is well-established in particle physics experiments such as neutrino physics and direct dark matter searches.
Although no LArTPC has as yet been developed for gamma-ray imaging spectroscopy, LArTPCs offer potential to work as a Compton camera with an unprecedentedly large effective area thanks to their intrinsic scalability.
In 2023, a three-channel LArTPC was successfully operated as an engineering balloon flight (eGRAMS) \citep{Nakajima2024}, and another flight experiment with a large LArTPC will be performed in 2026 (pGRAMS).

While GRAMS aims to deploy a LArTPC with a scale larger than \SI{1}{m}, we have also started a feasibility study of compact LArTPCs for use as Compton cameras. For this purpose, we developed a compact LArTPC, named NanoGRAMS \citep{Takashima2024a}.
In this work, we present the design of NanoGRAMS and evaluate its performance using a gamma-ray source.
This paper is organized as follows.
In Section~\ref{sec:detector_design}, we describe the detector design, including the low-noise scintillation and electron readout systems.
In Section~\ref{sec:gammaray_experiment}, we present the experimental setup and the results of laboratory measurements using a radioactive source to evaluate the performance of NanoGRAMS.
In Section~\ref{sec:discussion}, we discuss the analysis results of the gamma-ray measurements and outline future plans to improve the imaging performance.
Finally, in Section~\ref{sec:conclusion}, we summarize the conclusions of this study.

\section{Detector Design}
\label{sec:detector_design}
\subsection{Overview}
\begin{figure}[htbp]
    \centering
    \includegraphics[width=\linewidth]{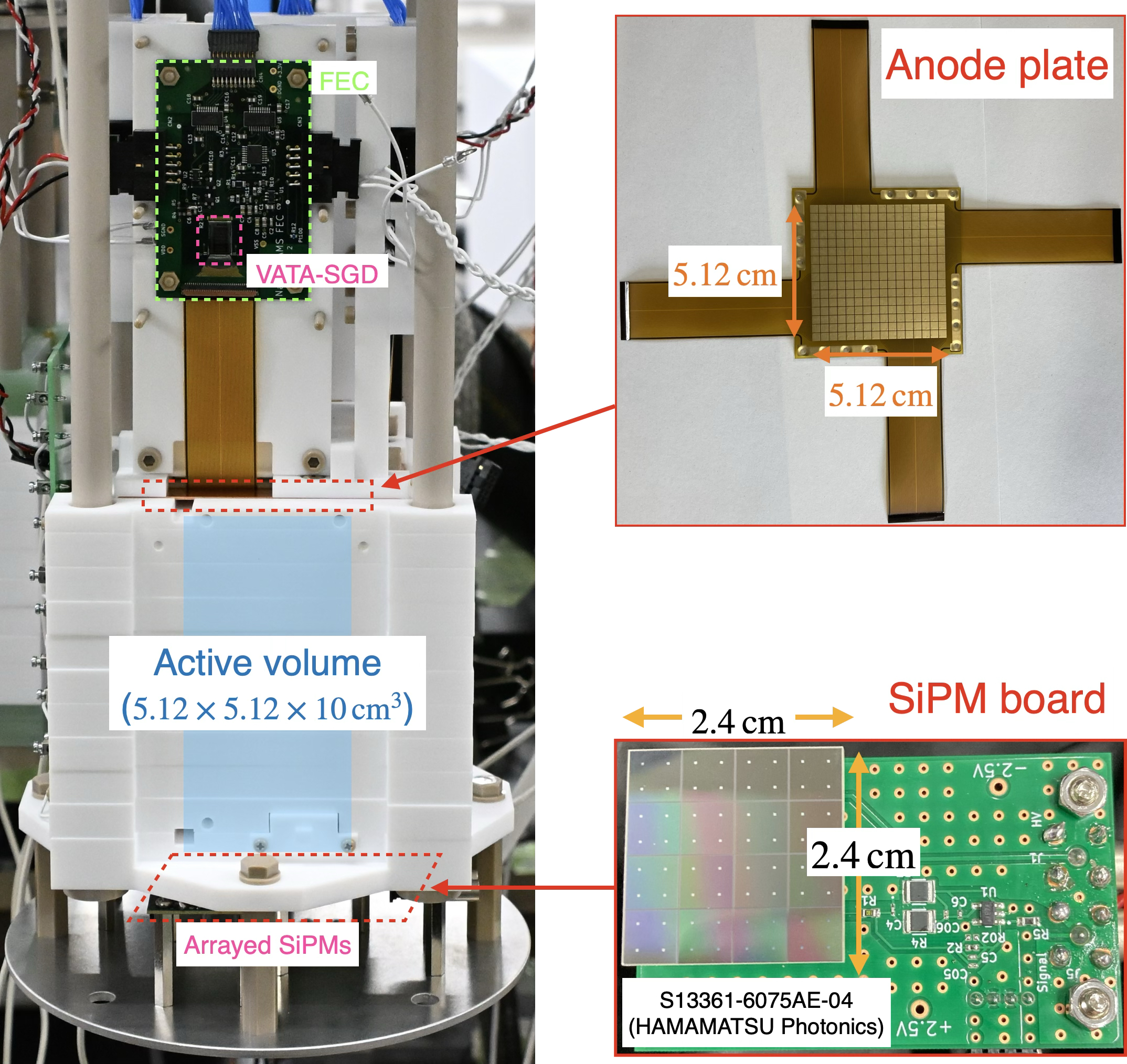}
    \caption{Photographs of NanoGRAMS and its components. During gamma-ray measurements, the front-end circuits are covered with PTFE sheets to maintain the operating temperature of the VATA-SGD ASICs above \SI{-40}{\degreeCelsius} for stable operation of their analog circuitry.}
    \label{fig:nanograms}
\end{figure}

NanoGRAMS has an active volume of $5.12 \times 5.12 \times 10~\mathrm{cm^3}$ (Fig.~\ref{fig:nanograms}). 
The detector structure is primarily made of polytetrafluoroethylene (PTFE) and polyether ether ketone (PEEK) to minimize outgassing and maintain argon purity. 
An anode plate, fabricated as a flexible printed circuit board, is mounted at the top of the active volume (upper right panel of Fig.~\ref{fig:nanograms}). 
At the bottom, a quartz glass plate coated with indium tin oxide (ITO) on one side is installed.
Because ITO is transparent to visible light and electrically conductive, the quartz glass plate serves as a transparent cathode when a negative voltage is applied to the ITO surface.
A uniform electric field along the drift direction is established not only by the cathode and anode but also by copper field-shaping electrodes embedded in the detector wall at \SI{1}{cm} intervals.

\begin{figure}[htbp]
    \centering
    \includegraphics[width=\linewidth]{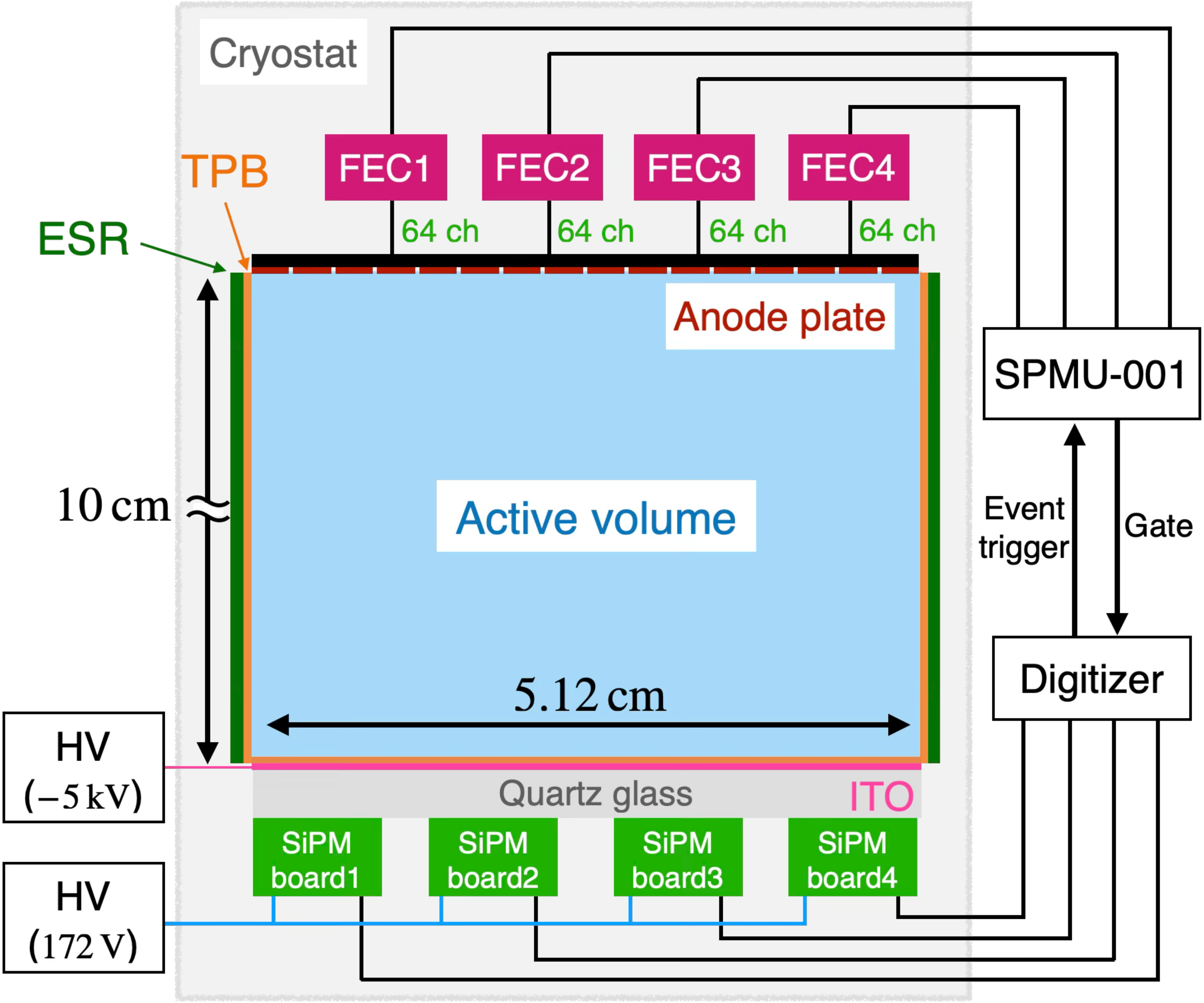}
    \caption{Schematics of the NanoGRAMS data acquisition system.}
    \label{fig:system}
\end{figure}

In the MeV energy range, gamma rays interact with liquid argon predominantly through photoabsorption and Compton scattering, producing localized energy depositions (hereafter referred to as hits). 
The LArTPC measures both the deposited energy and the three-dimensional position of each hit by detecting scintillation light at \SI{128}{nm} and ionization electrons generated in the liquid argon.
The prompt scintillation light is detected by four SiPM boards.
Each SiPM board is equipped with an array of 16 SiPMs ($6 \times 6\,\mathrm{mm^2}$ each) and has a single output port.
The four output signals are digitized by a 1~GS/s waveform digitizer (APU 8108-14, TechnoAP Co., Ltd.), which also provides the event trigger (see Fig.~\ref{fig:system}).
Ionization electrons drift under the applied electric field toward the anode, where they induce signals on \SI{3.2}{mm}-pitch pixel electrodes. 
These signals are amplified and digitized by four front end cards (FECs).

When one or more output waveforms from the SiPM boards exceed the trigger threshold, the digitizer issues an external trigger to the FPGA-based data acquisition board (SPMU-001), manufactured by Shimafuji Electric Incorporated.
The trigger threshold is configurable using the digitizer-control software and is usually set as low as possible without being triggered by dark signals
The SPMU-001 controls analog ASICs, called VATA-SGD, located on each FEC.
The SPMU-001 also receives self-trigger signals issued by the FECs and records the elapsed time from the digitizer trigger to each self-trigger (see Section \ref{sec:electron_readout} for details).
The drift time derived from the clock count determines the depth (drift coordinate; $z$) of each hit, while the pixel channel ID provides the transverse position; $x,y$. 
The deposited energy is reconstructed from the digitized charge amplitude.

\subsection{Photon detection system}
The scintillation light detection system consists of an array-type SiPM, or S13361-6075AE-04 manufactured by Hamamatsu Photonics and a transimpedance amplifier (TIA).
The TIA design is based on a preamp originally developed for direct dark matter searches \citep{DIncecco2018}. 
This design provides a fast response and stable operation at liquid argon temperature (\SI{87}{K}).
With appropriate modifications of passive components, the TIA was implemented on a printed circuit board on which the SiPM array is directly mounted. 
\begin{figure}[htbp]
\centering
\includegraphics[width=\linewidth]{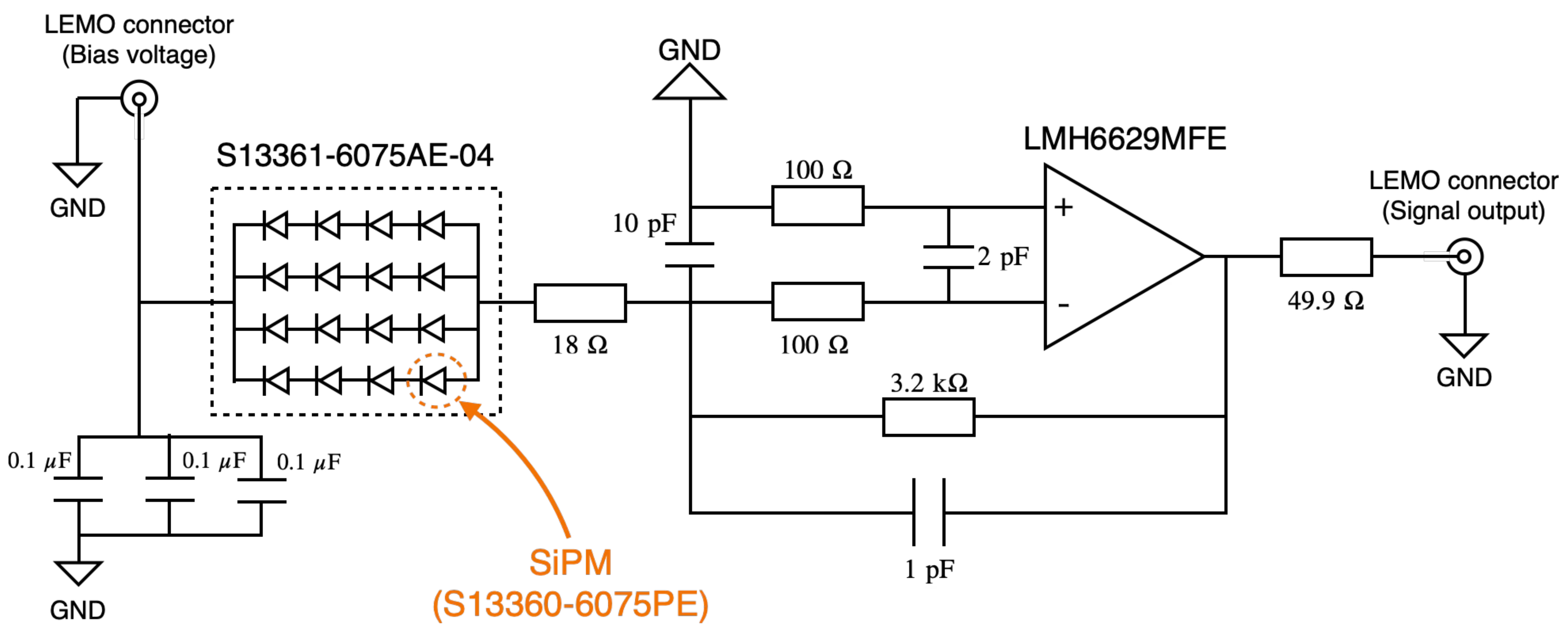}
\caption{Circuit configuration of the SiPM array and TIA.}
\label{fig:sipm_circuit}
\end{figure}

Outputs of each SiPM are combined into a single readout line to reduce the number of required waveform digitizer channels.
As shown in Fig.~\ref{fig:sipm_circuit}, four SiPMs are connected in series to form one unit, and multiple such units are connected in parallel. 
This configuration suppresses gain reduction and mitigates capacitance increase, thereby preventing an increase in decay time and associated dead time.
Dark current measurements were performed in liquid nitrogen (\SI{77}{K}). 
When \SI{176}{V} was applied to the SiPM board, corresponding to a bias voltage of \SI{44}{V} per SiPM, the gain and signal decay time were measured to be $1\times 10^6$ and \SI{86}{ns}, respectively.
No significant differences in gain or signal decay time were observed among the SiPM boards.

Scintillation light, which is originally in the vacuum ultraviolet band, is converted into visible light ($\sim 420\,\mathrm{nm}$) by the wavelength shifter, tetraphenylbutadiene (TPB), before arriving at the SiPMs.
TPB is deposited on enhanced specular reflectors (ESR) attached on the inner wall of the active volume and on the ITO-coated side of the transparent cathode.

\subsection{Electron detection system}
\label{sec:electron_readout}
The electron detection system consists of an anode pixel board, FECs, and SPMU-001.
The anode pixel board implements $16\times 16$ pixels and is connected to four FECs via four 64-line cables (see the upper right panel of Fig.~\ref{fig:nanograms}).
The VATA-SGDs on the FECs were originally developed for the Soft Gamma-ray Detector (SGD) onboard the Hitomi satellite \citep{Watanabe2014}.
It provides 64 input channels with low noise ($\sim 180~e^-$ for a \SI{6}{pF} input load).

The current signal from each input channel is fed into a charge-sensitive preamplifier (CSA), and the resulting voltage signal is split and sent to two shapers with fast and slow shaping times, referred to as the TA and VA parts, respectively.
When one of the TA outputs exceeds a threshold programmable by the SPMU-001, the ASIC issues a trigger signal to the SPMU-001.
The SPMU-001 then sends a sample-and-hold signal back to the ASIC after a fixed time delay, which can be adjusted.
When the sample-and-hold signal arrives, the all the VA outputs are sampled, held, and digitized in the ASIC.
Then, the digital values are read out by the SPMU-001.

Since the VATA-SGD is not designed for operation at liquid argon temperature, we attached a heater to the back of each FEC.
The temperature of each ASIC is estimated from the variation in the heater resistance and from a Pt100 resistance temperature detector attached next to the ASIC.
By controlling the heater, we maintain the ASIC temperature above \SI{-40}{\degreeCelsius}.
The ASIC gain correction for temperature variations is performed by injecting test pulses and acquiring calibration data during liquid argon experiments.

\section{Gamma-ray measurement}
\label{sec:gammaray_experiment}
In this section, we present an experiment to demonstrate performance of gamma-ray measurements with a radioactive isotope.

\subsection{Experimental setup}
\begin{figure}[htbp]
    \centering
    \includegraphics[width=\linewidth]{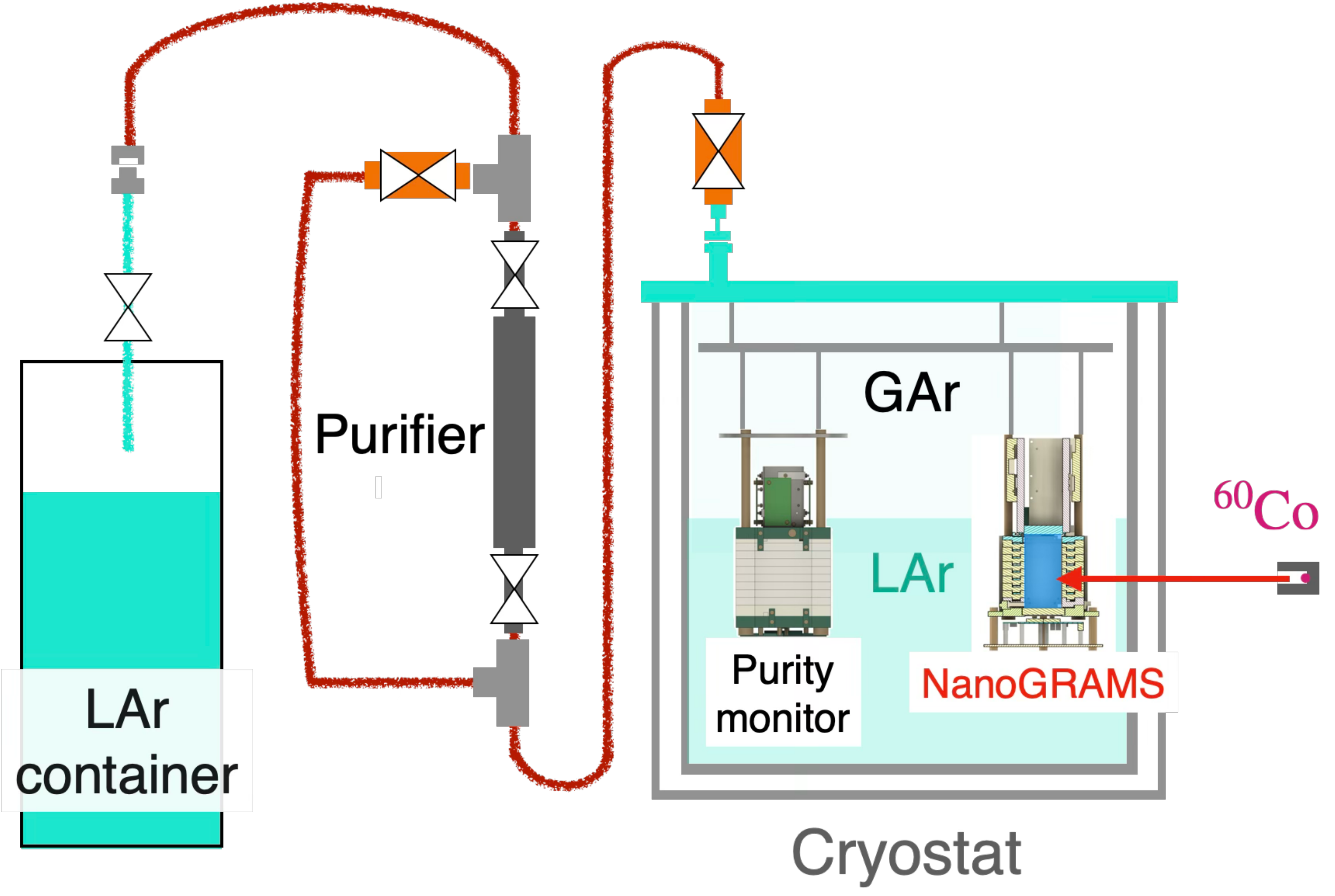}
    \caption{Experiment setup for the gamma-ray measurement.}
    \label{fig:cryostat_system}
\end{figure}

Figure~\ref{fig:cryostat_system} illustrates the gamma-ray measurement system.
NanoGRAMS is suspended inside a \SI{144}{L} stainless-steel cryostat filled with liquid argon.
The liquid level is adjusted to lie between the FECs and the anode plate to ensure stable operation of the readout electronics.
A $^{60}\mathrm{Co}$ source, emitting gamma rays of 1173 and \SI{1332}{keV}, was positioned outside the cryostat.
A lead collimator is placed in front of the source to suppress detection of scattered gamma rays.

Ionization electrons in liquid argon are susceptible to capture by electronegative impurities such as oxygen and water, leading to charge loss and degradation of energy resolution. 
The maximum drift time in NanoGRAMS is approximately \SI{70}{\mu s} at \SI{500}{V/cm}.
Therefore, the electron lifetime should be longer than \SI{100}{\mu s} to suppress charge loss during drift.
From a practical point of view, the concentration of such impurities is inversely proportional to the electron lifetime ($\tau_\mathrm{e}$). 
For oxygen contamination under zero electric field, the electron lifetime is approximately given by $\tau_\mathrm{e} \sim 300\,\mathrm{\mu s} \, ([\mathrm{O_2}]/\mathrm{ppb})^{-1}$, where $[\mathrm{O_2}]$ is the oxygen concentration \citep{Bakale1976,Acciarri2010}. 
This implies that impurity levels at few ppb level are required to allow electrons to drift without significant attenuation. 

To minimize contamination, the cryostat was evacuated to approximately \SI{1e-4}{Pa} prior to filling, in order to reduce outgassing from internal components.
Furthermore, liquid argon was passed through a hand-made purifier made from molecular sieve (Resonac Universal Corporation 4A XH-5 8$\times$12) and reduced copper (BASF PuriStar R3-11G T5$\times$3) before the filling to remove residual oxygen and water \citep{Curioni2009}, as shown in Fig.~\ref{fig:cryostat_system}.
A purity monitor was also installed adjacent to NanoGRAMS.
The monitor has the same sensitive volume ($5.12 \times 5.12 \times 10~\mathrm{cm^3}$) and a pixelated anode structure identical to that of NanoGRAMS.
Signals from its 64 pixel channels are combined section-wise, amplified by a CSA, and digitized.
During gamma-ray measurements, cosmic-ray muons traversing the full drift length of the purity monitor were recorded as full-drift events.
The full-drift event rate was one event every few seconds, which is consistent with the typical cosmic-ray muon rate in our laboratory on the fifth floor of a six-story building, considering the geometrical area of the purity monitor ($\sim$ \SI{25}{cm^2}).
Other backgrounds originating from radioactive decays in detector materials are not expected to contribute to the full-drift events because alpha and beta rays from such decays have typical energies of \SI{5}{MeV} and \SI{1}{MeV}, respectively.
These particles have ranges of less than \SI{1}{mm} in liquid argon and therefore produce much sharper waveforms in the CSA outputs.
An applied voltage of \SI{-1}{kV}, corresponding to an electric field of \SI{100}{V/cm}, was used for the purity monitor, while \SI{-5}{kV} was applied to the cathode of NanoGRAMS.
For the \SI{10}{cm} drift length of the purity monitor under a \SI{-1}{kV} bias voltage, the maximum drift time is approximately \SI{200}{\mu s}.
By analyzing the waveform amplitude and its dependence on drift time, we confirmed that no significant electron attenuation occurred during the measurements.
This indicates that the impurity concentration was below the ppb level, assuming oxygen contamination.

\subsection{Acquisition and calibration of gamma-ray data}
\begin{figure*}[htbp]
    \centering
    \includegraphics[width=\linewidth]{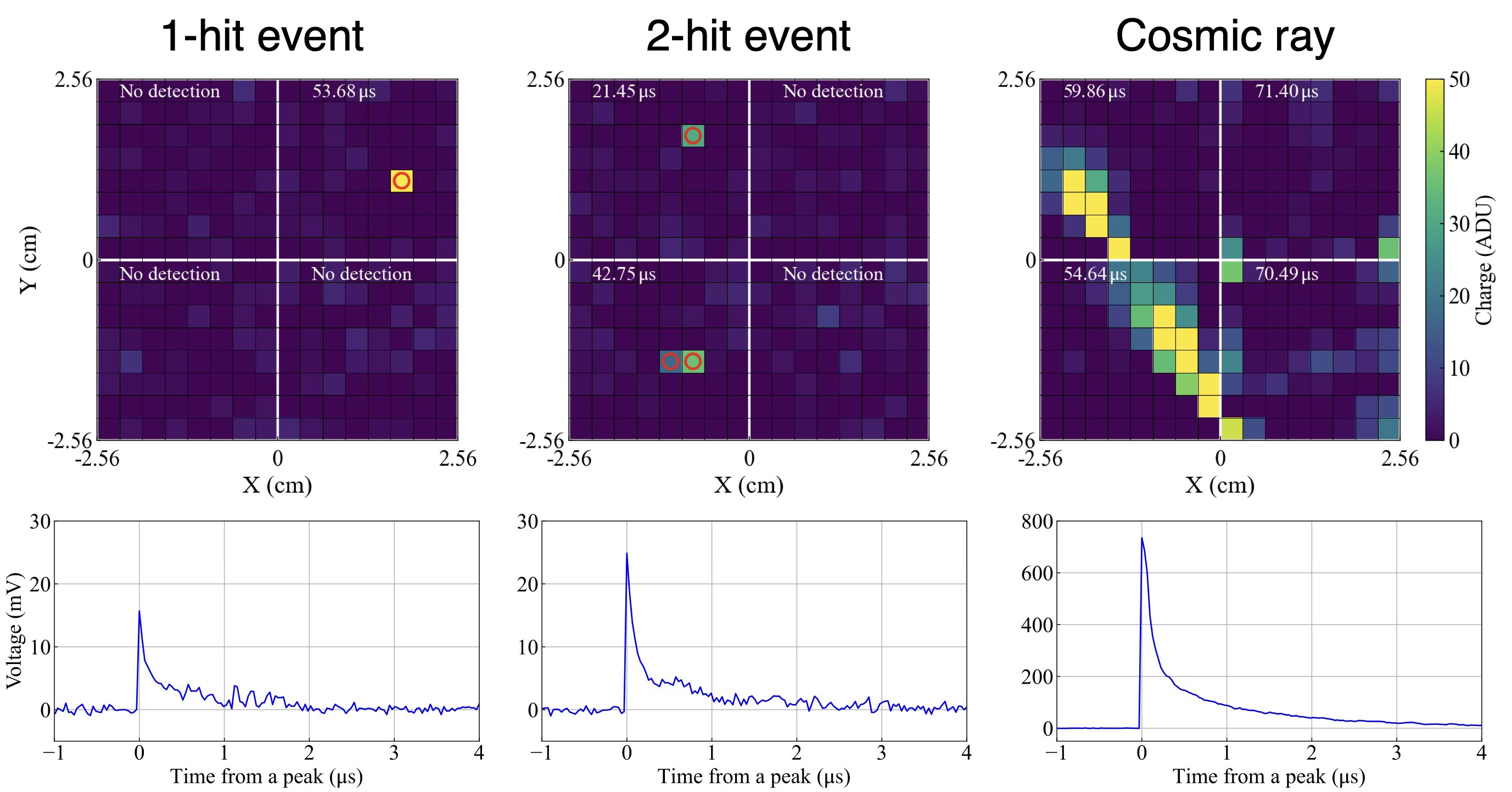}
    \caption{Representative event topologies recorded by NanoGRAMS. The top and bottom panels show electron signals induced on the anode plane and the corresponding scintillation light waveforms, respectively. The waveforms are averaged over the four SiPM boards. Left: a single-hit gamma-ray event showing a localized charge cluster confined within one pixel. Center: a 2-hit gamma-ray event with two spatially separated interaction points. Right: a cosmic-ray muon event forming an extended track across multiple anode sections. The times indicated in each quadrant represent the interval measured in the FPGA between the event trigger from the digitizer and the self-trigger signal of each VATA-SGD.}
    \label{fig:sample_event}
\end{figure*}

Light-triggered events were selected using the SiPM-based scintillation signal as the primary trigger. 
The common-mode noise of each VATA-SGD is estimated from the sampled-and-held CSA outputs of all 64 channels.
In the present work, we do not apply a depth correction because the pixel size is much smaller than the drift length, allowing the so-called small-pixel effect to suppress the depth dependence of the induced charge.
Figure~\ref{fig:sample_event} shows representative examples of electron signals induced on the anode plane and the corresponding scintillation light waveforms for 1-hit gamma-ray, 2-hit gamma-ray, and cosmic-ray events.
Because the pixel pitch of \SI{3.2}{mm} is larger than the typical extent of an electron track in liquid argon for MeV-scale Compton interactions, most gamma-ray hits are confined to one or two pixels. 
In contrast, cosmic-ray muons produce extended tracks spanning multiple pixels and anode sections, which are clearly distinguishable from gamma-ray topologies.
The time values indicated in each quadrant represent the time measured in the FPGA between the event trigger issued by the digitizer and the self-trigger timing of each VATA-SGD. 
The sum of this FPGA timing and the digitizer response time corresponds to the electron drift time, providing information on the interaction depth along the drift direction.
Scintillation light in liquid argon consists of fast and slow decay components ($\tau_\mathrm{f} \sim 7\,\mathrm{ns}$ and $\tau_\mathrm{s} \sim 1.6\,\mathrm{\mu s}$) \citep{Hitachi1983}. 
The ratio of these components depends on the particle type, and their separation is important for GRAMS to reject background events such as neutron-induced multiple scattering, which can mimic gamma-ray signals. 
All three event types demonstrate the capability to resolve both fast and slow components using our SiPM boards.

Full absorption of 1173 and \SI{1332}{keV} gamma rays in liquid argon is rare; therefore, conventional peak-based calibration is not applicable. 
We calibrated the detector response by converting 10-bit digitized charge values to deposited energy (keV) through a two-step procedure.

First, the electronic response was calibrated using the built-in DAC of the VATA-SGD. 
Injected charges were applied to establish the charge-to-ADC conversion. 
For a localized energy deposition ($E$), the collected charge ($Q$) is expressed as
\begin{equation}\label{eq:collected_charge}
    Q = \frac{eE}{w_i} R(E,\mathcal{E}),    
\end{equation}
where $e$ is the elementary charge, $\mathcal{E}$ is the applied electric field, $w_i=23.6\,\mathrm{eV}$ is the effective ionization energy in liquid argon \citep{Miyajima1974}. 
$R(E,\mathcal{E})$ is the recombination probability of electron–ion pairs, which is a function of the energy and the applied electric field.
In this paper, we deal only with data acquired for $\mathcal{E}=500\,\mathrm{V/cm}$.
Following \cite{Segreto2024}, the recombination probability $R(E,\mathcal{E})$ is modeled as
\begin{equation}\label{eq:recombination_probability}
    R(E,\mathcal{E}) = \frac{\mathcal{E}}{\mathcal{E} + \frac{k\beta}{w_i}}\left(1 + \frac{k\gamma}{\mathcal{E}}\right)\left[1 - \frac{\ln(1+z(E,\mathcal{E}))}{z(E,\mathcal{E})}\right],
\end{equation}
with
\begin{equation}\label{eq:zdef}
    z(E,\mathcal{E}) = \frac{\mathcal{E} + \frac{k}{w_i}\beta}{\frac{k}{w_i}\alpha}E,
\end{equation}
where $\alpha = 0.227\,\mathrm{MeV^2/cm}$, $\beta = 1.7\,\mathrm{MeV/cm}$, $k = 3.7\,\mathrm{mV}$, and $\gamma = 2.8\,\mathrm{\mu m^{-1}}$ \citep{Aprile1987}. 
Details of the model are provided in Appendix~A.

Second, to account for residual effects not included in the electronic calibration and recombination model, we performed a Monte Carlo simulation based on the Geant4 toolkit \citep{Allison2006,Agostinelli2003} via the ComptonSoft framework \citep{Odaka2010}. 
Gamma-ray interactions from the $^{60}\mathrm{Co}$ source were simulated under the same geometrical configuration as in the experiment.
The geometry model included the active liquid argon volume, PTFE walls, and the stainless cryostat.
To incorporate the effect of the detector energy resolution, simulated spectra were generated assuming multiple energy resolutions. 
For each assumed resolution, a scale factor $s$ was introduced, and the agreement between the measured and simulated spectra was evaluated. 
For a given energy resolution $\Delta E/E$ and scale factor $s$, the $\chi^2$ value between the normalized simulated spectrum $N_{\mathrm{sim}}(i)$ and the scaled measured spectrum $N_{\mathrm{data}}(i; s)$ was calculated as
\begin{equation}
    \chi^2(s)=\sum_i \frac{\left( N_\mathrm{data}(i;s)-N_\mathrm{sim}(i) \right)^2}{\sigma_i^2},
\end{equation}
where $i$ is an index of energy bin and the statistical uncertainty was defined as $\sigma_i = \sqrt{N_{\mathrm{data}}(i; s)}$.
The reduced chi-square $\chi^2/\nu$, where $\nu$ is the number of degrees of freedom, was used to evaluate the goodness of fit. 
The scale factor $s$ that yields $\chi^2/\nu$ closest to unity was adopted as the energy-scale correction factor.

\begin{figure}[htbp]
    \centering
    \includegraphics[width=\linewidth]{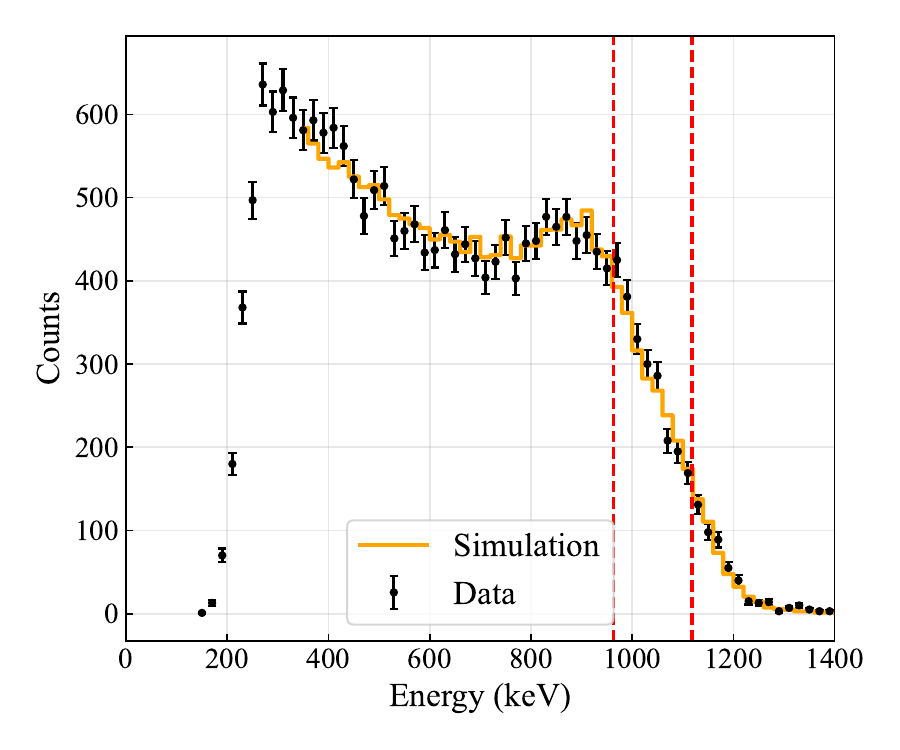}
    \caption{Reconstructed energy spectrum of $^{60}\mathrm{Co}$ gamma rays (black) obtained from the electron signals compared with the best-fit Geant4 simulation (orange). The red dashed lines indicate the expected Compton-edge positions at 963 and \SI{1118}{keV}. The simulation reproduces both the spectral shape and the Compton-edge features above \SI{350}{keV}, which was used as the normalization region.}
    \label{fig:spectrum}
\end{figure}

Figure~\ref{fig:spectrum} shows the reconstructed energy spectrum of $^{60}\mathrm{Co}$ obtained from the electron signals together with the best-fit simulated spectrum. 
The lower-energy region is affected by the trigger thresholds and event-selection conditions.
Therefore, the energy region above \SI{350}{keV} was used for spectrum fitting and shape comparison.
The red dashed lines indicate the expected Compton-edge energies at 963 and \SI{1118}{keV} for the 1173 and \SI{1332}{keV} gamma rays, respectively. 
The measured spectrum reproduces these Compton edges and agrees well with the simulation over the fitted energy range. 
The best-fit energy resolution is 3.6\% at an energy deposition of \SI{2.5}{MeV}. 
The contribution of electronic noise from the readout system is evaluated using the ASIC charge injection function. 
The noise level is estimated from pedestal fluctuations measured in selected pixels using triggered events in which no gamma-ray signal is observed in those pixels. 
The standard deviation of these fluctuations corresponds to an equivalent noise charge (ENC) of 130--160$~e^-$.
Further discussion of the achievable energy resolution is presented in Section~\ref{sec:discussion}.

\subsection{Back-projection image analysis}
For 2-hit events, Compton imaging can be performed if the first and second interactions are correctly identified and the incident gamma-ray energy is known. 
In this study, the initial gamma-ray energy was assumed from the known emission lines of the radioisotope source, and a back-projection analysis was applied to 2-hit events.
The interaction order was inferred using energy–momentum conservation and a figure-of-merit criterion.
For each 2-hit event, two possible interaction sequences were considered. 
For a given ordering, the Compton scattering angle $\theta$ was calculated using
\begin{equation}
    \cos\theta = 1 - m_\mathrm{e} c^2
    \left(
    \frac{1}{E_0 - E_1} - \frac{1}{E_0}
    \right),
\end{equation}
where $E_0$ is the incident gamma-ray energy and $E_1$ is the energy deposited at the first interaction point. 
If the right-hand side exceeds the physical range ($|\cos\theta| > 1$), the corresponding interaction order is rejected.
When both order of interaction satisfy $|\cos\theta| \leq 1$, the geometric scattering angle $\theta_\mathrm{G}$ is calculated from the reconstructed hit positions. 
Here, $\theta_\mathrm{G}$ is defined as the angle between (i) the vector from the first interaction point to the known source position and (ii) the vector from the second to the first interaction point. 
A sequence that minimizes $|\theta - \theta_\mathrm{G}|$ is selected as the most probable ordering.

\begin{figure}[htbp]
\centering
\includegraphics[width=\linewidth]{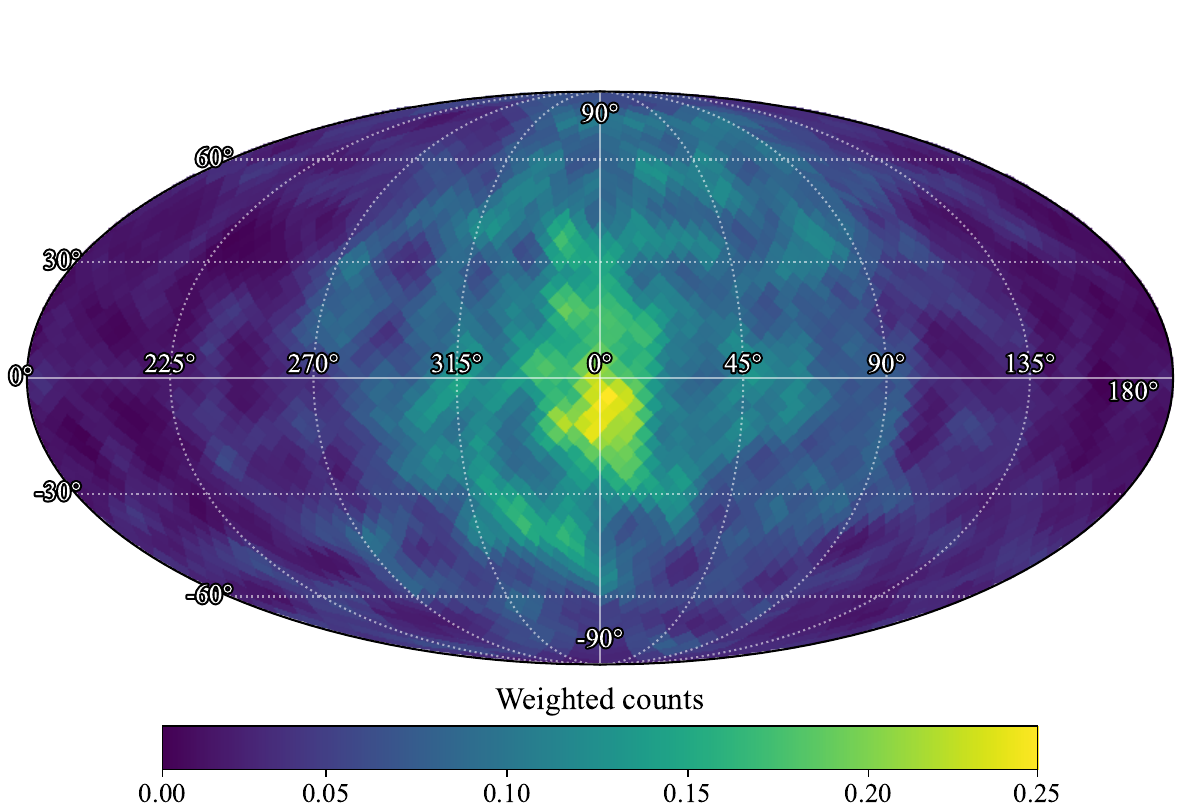}
\caption{Back-projection image reconstructed from the 2-hit $^{60}\mathrm{Co}$ events. The Compton cones were projected onto a $4\pi$ sphere and binned using the HEALPix equal-area pixelization scheme. The map is shown in a Mollweide projection.}
\label{fig:backprojection}
\end{figure}

Using the selected 2-hit events, Compton cones were reconstructed and projected onto a $4\pi$ celestial sphere centered on the detector.
The sphere was pixelized using the Hierarchical Equal Area isoLatitude Pixelization (HEALPix) scheme \citep{Gorski2005}, which provides equal-area pixels on a sphere.
The HEALPix resolution parameter, \textit{nside}, was set to 16, corresponding to 3072 pixels over the full sky.
The solid angle of each pixel is equivalent to that of a circle with an angular radius of approximately $2.1^\circ$.
Each projected Compton cone was smoothed with a Gaussian kernel with a standard deviation of $3^\circ$, which is comparable to a typical angular resolution of Compton cameras.
Figure~\ref{fig:backprojection} shows the resulting back-projection image.
A concentration of reconstructed intersections is observed near the true source position, $(X,Y)=(0^\circ,0^\circ)$.
The observed displacement of the brightest point from the true position, $\Delta Y \sim 5^\circ$, can be explained by the uncertainty in the relative positioning between NanoGRAMS and the source.

\section{Discussion}
\label{sec:discussion}
We have demonstrated the performance of NanoGRAMS as a prototype Compton camera using liquid argon.
In this section, we discuss possible improvements to enhance the performance of LArTPC-based Compton cameras.
The angular resolution of a Compton camera is fundamentally limited by the energy and spatial resolutions of individual gamma-ray interactions. 
Improving the energy resolution requires a more accurate reconstruction of the deposited energy, for which enhanced scintillation-light collection is essential in addition to ionization charge readout. 

The nominal noise performance of the readout electronics ($5\,\mathrm{keV}$) \citep{Aramaki2020} is satisfied in this system. 
For a gamma ray depositing \SI{1}{MeV} under an electric field of \SI{500}{V/cm}, the expected number of collected electrons is $2.85 \times 10^4\,e^-$, as calculated using Eq.~(\ref{eq:collected_charge}). 
In this case, the measured electronic noise of $140\,e^-$ in NanoGRAMS corresponds to $4.9\,\mathrm{keV}$, which is consistent with the nominal value. 
Therefore, the current energy resolution is dominated by photon statistics and light-collection efficiency. 

The spatial resolution is constrained by the anode segmentation with a pixel pitch of \SI{3.2}{mm}. 
Although a finer pitch would significantly increase the number of readout channels, it would improve position reconstruction. 
The impact of these improvements on Compton imaging performance has been quantitatively studied in \citep{Arai2026}.

The current front-end ASIC (VATA-SGD) was not designed for operation at liquid argon temperature. 
Stable operation of its analog circuitry requires local heating, which increases power consumption and complicates thermal management. 
The development of a cryogenic-compatible front-end ASIC would improve system reliability, simplify detector integration, and enable more efficient large-scale LArTPC-based Compton cameras.

\section{Conclusions}
\label{sec:conclusion}
We have developed the NanoGRAMS detector, with dimensions of $5.12 \times 5.12 \times 10\,\mathrm{cm^3}$, for a concept study of GRAMS. 
Since imaging spectroscopy with a Compton camera requires detectors with high energy and position resolution, 
NanoGRAMS is designed with a low-noise light and electron readout system using arrayed SiPM circuits and VATA-SGDs.
Using NanoGRAMS, we successfully measured gamma-ray 1-hit and 2-hit events from a $^{60}\mathrm{Co}$ source (1173 and \SI{1332}{keV}).
For the analysis of 2-hit events, we identified the first and second interactions that occur sequentially. 
The reconstructed back-projection image is consistent with the actual position of the $^{60}\mathrm{Co}$ source.
These results demonstrate the feasibility of NanoGRAMS as a Compton camera for MeV gamma-ray imaging. 
Planned upgrades, including improvements to the light and electron readout systems, are expected to further enhance the angular and energy resolutions.

\section*{Acknowledgement}
The authors thank anonymous reviewers for insightful comments which improve the manuscript.
This work and HO have been supported by Toray Science and Technology Grant No.~20–6104 (Toray Science Foundation) by JSPS KAKENHI Grant-in-Aid for Scientific Research on Innovative Areas Numbers 18H05861, 19H01906, 19H05185, 22H00128, 22K18277(HO), 20K22355(HY), and 22KJ0811(ST).
This work is also supported by the NASA APRA grant No. 22-APRA22-0128 (80NSSC23K1661).
The cryostat was manufactured by ARIOS INC.

\section*{Declaration of generative AI and AI-assisted technologies in the manuscript preparation process}
During the preparation of this work the authors used ChatGPT 5.5 in order to improve language and readability. 
After using this tool, the authors reviewed and edited the content as needed and take full responsibility for the content of the published article.


{}

\appendix
\section{Recombination probability}
\label{app1}

In this section, we derive Eq.~(\ref{eq:recombination_probability}) following Ref.~\cite{Segreto2024}.
Let $dq_i$ be an infinitesimal ionization charge created promptly after a gamma-ray interaction. 
A fraction of these electrons avoids recombination with probability 
$\overline{P}(\mathcal{E},\, dq_i/dx,\, Q_i)$, 
which depends on the electric field $\mathcal{E}$, the stopping power $dq_i/dx$, and the total initial ionization charge $Q_i$.
The infinitesimal charge collected at the anode, $dq$, is written as
\begin{equation}
    dq = dq_i \times \overline{P}(\mathcal{E},\, dq_i/dx),
\end{equation}
where $dq_i = dE / w_i$, and $w_i = 23.6\,\mathrm{eV}$ as described in the main text.
By integrating over the total produced charge, the fraction of electrons extracted to the anode, $\overline{R}(E,\mathcal{E})$, is expressed as
\begin{equation}
    \overline{R}(E,\mathcal{E}) 
    = \frac{1}{Q_i} \int_0^{Q_i} dq 
    = \frac{1}{Q_i} \int_0^{Q_i} dq_i \times \overline{P}(\mathcal{E},\, dq_i/dx).
\end{equation}
For stopping protons and muons, the extraction probability can be parameterized as
\begin{equation}
    \overline{P}(\mathcal{E},\, dq_i/dx) 
    = \frac{\mathcal{E}}{k \frac{dq_i}{dx} + \mathcal{E}},
\end{equation}
where $k$ is an empirical constant.
The stopping power is approximated as
\begin{equation}
    \frac{dE}{dx} \simeq \frac{\alpha}{E} + \beta,
\end{equation}
where $E$ is the kinetic energy of the beta particle, 
$\alpha = 0.227\,\mathrm{MeV^2/cm}$, 
and $\beta = 1.7\,\mathrm{MeV/cm}$.
Substituting these expressions into the definition of $\overline{R}(E,\mathcal{E})$, 
the extraction fraction for a total deposited kinetic energy $E$ becomes
\begin{align}
    \overline{R}(E,\mathcal{E})
    &= \frac{1}{Q_i}\int_0^{Q_i} dq_i \times \overline{P}(\mathcal{E},\,dq_i/dx) \\
    &= \frac{1}{E_\mathrm{kin}} 
       \int_0^{E_\mathrm{kin}} 
       dE \,
       \frac{\mathcal{E}}
       {\frac{k}{w_i}\left(\frac{\alpha}{E} + \beta\right) + \mathcal{E}} \\
    &= \frac{\mathcal{E}}{\mathcal{E} + \frac{k\beta}{w_i}}
       \left(1 - \frac{\ln(1+z(E,\mathcal{E}))}{z(E,\mathcal{E})}\right),
\end{align}
where the dimensionless parameter $z(E,\mathcal{E})$ is defined in Eq.~(\ref{eq:zdef}).

Experimental studies have shown that a fraction of electrons escapes recombination even in the absence of an external electric field (S. Kubota 1978, T. Doke 1990). 
Let $S(dq_i/dx)$ denote the fraction of electrons that survive recombination without an applied field. 
The effective extraction probability is then written as
\begin{equation}
    P_\mathrm{e} = \overline{P}(\mathcal{E},\, dq_i/dx)  + S(dq_i/dx) (1 - \overline{P}(\mathcal{E},\, dq_i/dx) ).
\end{equation}

By analogy with the parameterization of $\overline{P}$, the recombination escape fraction is assumed to take the form
\begin{equation}
    S(dq_i/dx) = \frac{1}{1 + \frac{dq_i/dx}{\gamma}} 
    \simeq \frac{\gamma}{dq_i/dx},
\end{equation}
where $\gamma$ is an experimentally determined parameter. 
The approximation is valid when $dq_i/dx \gg \gamma$.
Finally, the fraction of collected electrons, $R(E,\mathcal{E})$, becomes
\begin{equation}
    R(E,\mathcal{E}) = \frac{1}{Q_i}
       \int_0^{Q_i} P_\mathrm{e}\,dq_i = \left(1 + \frac{k\gamma}{\mathcal{E}} \right) \overline{R}(E,\mathcal{E}).
\end{equation}

\end{document}